\pdfoutput=1

\documentclass{article}

\usepackage{icrc2011}
\usepackage{amssymb}

\title{Prospects of GRB observations for CTA from a phenomenological model}

\newcommand{\etal}{\MakeLowercase{\textit{et al. }}} 
\shorttitle{Bouvier \etal CTA GRB estimation}

\authors{Aurelien Bouvier$^{1}$, Rudy Gilmore$^{2}$, Valerie Connaughton$^{3}$, Nepomuk Otte$^{1}$, Joel R. Primack$^{4}$, David A. Williams$^{1}$ }
\afiliations{$^1$Santa Cruz Institute for Particle Physics, University of California, Santa Cruz, CA 95064, USA\\ 
$^2$SISSA - {\it Astrophysics Sector} - Via Bonomea 265 - 34136, Trieste - Italy\\
$^3$National Space Science \& Technology Center, University of Alabama in Huntsville, Huntsville, AL 35805, USA \\
$^4$Department of Physics, University of California, Santa Cruz, CA 95064, USA}
\email{{\rm contact email: }apbouvie@ucsc.edu}

\abstract{Very high energy (VHE, i.e. $\gtrsim 10$ GeV) photons from Gamma-Ray Bursts (GRBs), as high as 90 GeV in rest frame energy, have been detected by the $Fermi$ Large Area Telescope (LAT). This provides hope for a high statistics GRB detection by a ground-based instrument in the VHE domain. We here report on our estimates of the expected GRB detection rate and $\gamma$-ray rate (in case of detection) for the next generation ground-based Imaging Air Cherenkov Telescopes (IACTs): the Cherenkov Telescope Array (CTA). Moreover, we investigated the effect critical design parameters of CTA have on our predictions and we compare the performance of various observing strategies for the array pointing. Our estimations are based on a phenomenological model which uses temporal and spectral information from GRBs detected by $Fermi$-LAT as well as other instruments operating at lower energy. While detection of VHE emission from GRBs has eluded ground-based instruments thus far, our results suggest it should be within reach of CTA with a rate between 0.35 and 1.6 GRBs/year depending on the characteristics of the true GRB population and the performance that CTA will eventually achieve.}

\begin{document}
\maketitle

\section{Introduction}

Detecting VHE emission from GRBs with ground-based telescopes would be a significant step forward not only for our comprehension of the underlying physics of these catastrophic events but also help us probe the evolving infrared-optical-ultraviolet extragalactic background light (EBL) at relatively high redshifts. Such VHE detection would most likely yield high photon statistics opening the door to detailed time-resolved spectroscopy in the VHE regime which would provide important clues on the emission processes, particle distributions, and internal absorption since current lack of statistics in this regime (only a handful of $\gamma$-ray photons were detected above 10 GeV by  $Fermi$-LAT) prevent a clear distinction between different spectral models (power-law, broken power-law, smooth curvature, exponential cutoff...). Furthermore, by probing the highest $\gamma$-ray energies emitted during prompt and afterglow emissions, VHE observations would set significant constraints on the particle acceleration mechanisms, the possible presence of very-high energy cosmic-rays and the bulk Lorentz factor of the jet.
Finally, by searching for signature of absorption through intervening IR-optical-UV light, important observational constraints on the EBL content could be derived.

GRBs are a prime objective of current IACTs (MAGIC, VERITAS, HESS) but attempts of detection have proved fruitless so far. The next generation instrument CTA\footnote{http://www.cta-observatory.org/} probably holds the best chances at succeeding in this endeavor thanks in particular to its order of magnitude higher effective area and few tens of GeVs energy threshold.
A serendipitous observation of a GRB onset in the Field-of-View (FoV) of an IACT is extremely unlikely ($\sim 1\%$ chance every year assuming a whole sky rate of $\sim 600$ GRBs/year and $\sim 4^{\circ}$ FoV). Therefore, Cherenkov telescopes rely on external GRB alerts in order to point  their telescope arrays toward a GRB candidate for VHE emission.
We now make a little tour of GRB-dedicated mission (to the best of our knowledge) that might be in orbit at the start of CTA operation currently scheduled for $\sim 2018$, and would therefore be capable of providing such external alerts:

\begin{itemize}
\item {\it Swift} \cite{gehrels09}: alert rate $\sim 95$ GRBs/year with extremely good localization ($\lesssim 10$'')
\item SVOM \cite{SVOM}: alert rate $\sim 70-90$ GRBs/year with similarly good localization; expected launch date $\sim~2015$
\item $Fermi$-GBM \cite{GBM}: alert rate $\sim 250$ GRBs/year with poor localization (several degrees)
\end{itemize}

Because the composition of the satellite fleet that will be in orbit at the time of CTA's first light remains quite uncertain, our approach is to provide independent estimates for the two vastly different type of GRB alerts that might be available: 1) {\it Swift} and/or SVOM, 2)~{\it Fermi}-GBM.

\section{Modeling the GRB population at VHE}

`Make things as simple as possible, but no simpler' is the spirit we tried to follow to build our phenomenological model. The lack of knowledge about GRB VHE emission naturally dominates the uncertainties in our predictions. We estimated the level of this uncertainty by considering GRB populations with different spectral and temporal properties in the VHE domain, a range of behavior that can be reasonably assumed to encompass the properties of the true GRB population.

\subsection{Temporal and spectral model}

The VHE light curve is considered flat during the prompt emission to which a duration drawn from the BATSE $T_{90}$ distribution is assigned. Following the prompt emission, an extended VHE emission was modeled with a temporal decay $t^{-1.5}$ similar to the decay measured on bright LAT GRBs \cite{ghisellini10}\footnote{temporal decay of $t^{-1.2}$ and $t^{-1.8}$ were also simulated without any significant change in the conclusions of our study.}.

Because the VHE spectrum is highly unknown and will strongly affect our predictions, we consider two different spectral models:

\begin{itemize}
\item `Bandex' model: a simple extrapolation of the Band function to VHE where we impose a maximum value of -2.0 to the high energy index.

\item `Fixed' model: a power-law component is added on top of the Band function with an index -2.0 and normalization chosen to fix the energy flux ratio between the LAT  ($100 \rm{ MeV}-300\rm{GeV}$) and BATSE ($50-300 \rm{ keV}$) to 10\%.
\end{itemize}

Band function parameters are all drawn from  the BATSE distributions. For {\it Swift} simulated bursts, a global fluence multiplier of 0.75 is applied to the BATSE fluence distribution to provide the best fit between BATSE and {\it Swift}/BAT fluences in the $15-150$ keV range

These two spectral scenarios are roughly consistent with the LAT detection rate of $\sim 10$ GRBs/year (`Bandex': $\sim~10$ GRBs/year; `Fixed': $\sim~20$ GRBs/year). We also note that `Bandex' and `Fixed' spectral model match the behavior observed on the bright LAT GRBs: 080916C and 090902B respectively.

Let us point out that internal spectral curvature is not considered mostly because of the large uncertainty this feature creates. Indeed, depending on the level of curvature assumed, the VHE signal could either completely disappear or be almost unaffected. As a consequence of this caveat, the actual CTA observations could be substantially lower than our predictions if strong internal curvature below $\sim 100$ GeV are common in GRB spectrum. However, we point out that we consider very unlikely for the CTA GRB detection rate to be significantly higher than our prediction. In that sense, our estimates can be seen as upper limits for the actual CTA GRB detection rate. 

\subsection{Redshift distribution}

Redshifts of our simulated bursts are drawn from the {\it Swift} population of $\sim 170$ GRBs with measured redshifts\footnote{http://heasarc.gsfc.nasa.gov/docs/swift/archive/grb\_table/}. GBM (SVOM) is (will be) sensitive to a slightly different redshift distribution but we consider this difference to be a second order effect in our predictions.

The EBL model used in our analysis is Gilmore \& Somerville 2011 \cite{Gilmore2011} approximately consistent with Gilmore et al. (2009), Franceschini et al. (2008), Finke et al. (2010), and Dominguez et al. (2010). We are currently investigating the effect that different EBL models would have on our estimates.

\section{CTA instrument response}

At the time of writing of this proceeding, the performance of the Cherenkov Telescope Array is still largely uncertain, in particular its low energy response which is crucial for GRB observations since  extragalactic events suffer strong EBL absorption in the VHE domain.
For this reason, we refrain from accurately modeling the CTA response but instead consider two CTA performances which we believe should encompass the final true CTA performance:

\begin{itemize}

\item CTA `baseline': 4 central Large Size Telescopes (LSTs) with energy threshold $E_{th} = 25$ GeV; 25 Medium Size Telescopes (MSTs).

\item CTA `optimistic': 4 LSTs with $E_{th} = 10$ GeV; 75 MSTs. Background rate artificially lowered by a factor of 3 to consider possible improvement in cosmic-ray rejection.

\end{itemize}

 \begin{figure}[!t]
  \vspace{5mm}
  \centering
  \includegraphics[width=3.in]{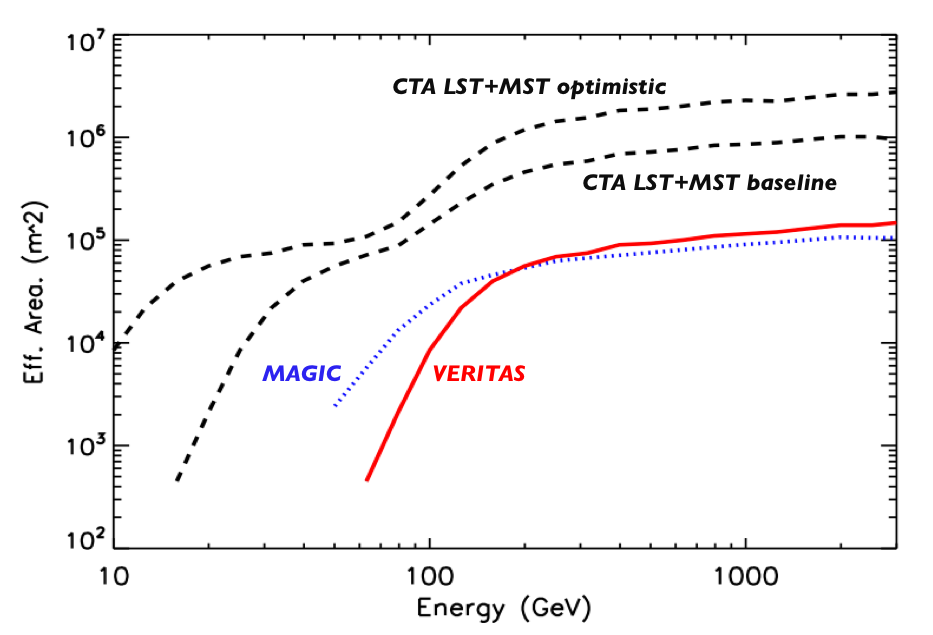}
  \caption{Effective areas for MAGIC, VERITAS and the two CTA performances we are using in our study: baseline and optimistic.}
  \label{EA}
 \end{figure}

Effective areas (see fig. \ref{EA}) and cosmic-ray rates for the LSTs and MSTs are derived through simple scaling of the well known VERITAS instrument performance and assuming a cosmic-ray spectrum of $E^{-2.7}$. Small Size Telescopes (SSTs) are not considered since they are mostly sensitive to $\gtrsim 1$~TeV showers. Zenith dependence of the energy threshold is assumed to be $E_{th} = E_{th,0} \times cos(\Theta_{Zenith})^{-3}$. 
As for the time delay it will take CTA to start observing the field of a GRB from the time it was actually triggered\footnote{this time includes the time for the GCN alert to be sent out to the ground and received on site, for the observers to authorize slewing of the telescopes and for the actual slewing time of the telescopes to reach the target field.}, we assume 60 seconds for the LSTs and 100 seconds for the MSTs and show in section \ref{design_param} the effect varying this parameter has on our predictions.
Finally our estimates are based on one Cherenkov array only. In case the CTA collaboration is able to build a second array (one in each hemisphere) with similar characteristics, our predicted rates would naturally increase by a factor of 2.

\section{Results}

\subsection{Detection and $\gamma$-ray rates}

\begin{figure}[!t]
  \vspace{5mm}
  \centering
  \includegraphics[width=3.in]{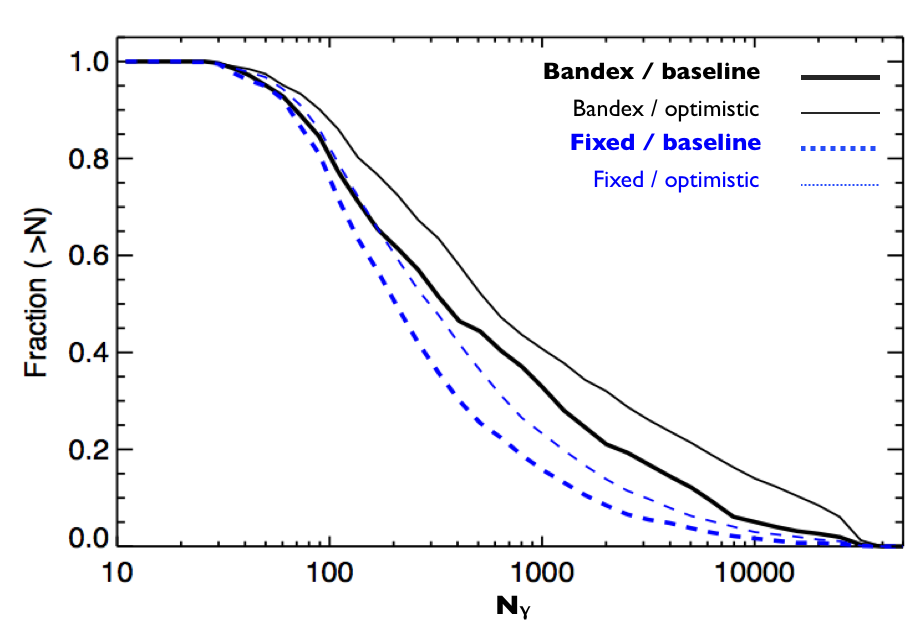}
  \caption{Number of $\gamma$-ray photons detected by CTA in case of detection}
  \label{stats}
 \end{figure}

For each of the GRB population models (`bandex' and `fixed') and each of the CTA performances (`baseline' and `optimistic'), we randomly simulated 10,000 GRBs in a solid angle $75^{\circ}$ around zenith. We then computed the detection efficiency for these bursts, i.e. the probability of detecting a GRB when the Cherenkov array was able to observe the GRB field (i.e. for a GRB that occurs during a moonless night with good weather and at least $15^{\circ}$ above the horizon). Assuming a {\it Swift} alert rate of 95 GRBs/year with a solar anti-bias factor of 1.4 \cite{Gilmore2010} and a $10\%$ duty cycle, table \ref{DR} provides our predictions for the yearly GRB detection rate for various IACT performances (MAGIC, CTA baseline, CTA optimistic), type of satellite alerts ({\it Swift}, GBM), and spectral model (`bandex', `fixed').
In case such detection is established, fig. \ref{stats} shows it will almost certainly yield high photon statistics with a median number of $\gamma$-rays expected around 200-600 depending on the assumptions on the GRB model and the CTA performance.

\begin{table}[t]
\begin{center}
\begin{tabular}{l|ccc}
\hline
 & {\it Swift}  &  {\it Swift} & GBM \\
 & `bandex'  & `fixed' & `fixed' \\
\hline
MAGIC  & 0.1 & 0.15 & -    \\
CTA baseline  & 0.35 & 0.6 & 0.7  \\
CTA optimistic  & 0.8 & 1.6 & 1.6 \\
\hline
\end{tabular}
\caption{GRB yearly detection rates for various IACT performance (MAGIC, CTA baseline, CTA optimistic), type of satellite alerts ({\it Swift}, GBM), spectral model (`bandex', `fixed'). Note that the GBM numbers are obtained when simulating the `orbit' mode observing strategy (see section \ref{GBM}).}
\label{DR}
\end{center}
\end{table}

The redshift probability distribution of detected bursts is naturally shifted from the {\it Swift} distribution towards lower redshift due to the absorption by the EBL. The median redshift of detected bursts is for example expected to be around redshift of 1.0.

\subsection{Effect of critical CTA design parameters}
\label{design_param}

\begin{figure}[!t]
  \vspace{5mm}
  \centering
  \includegraphics[width=3.in]{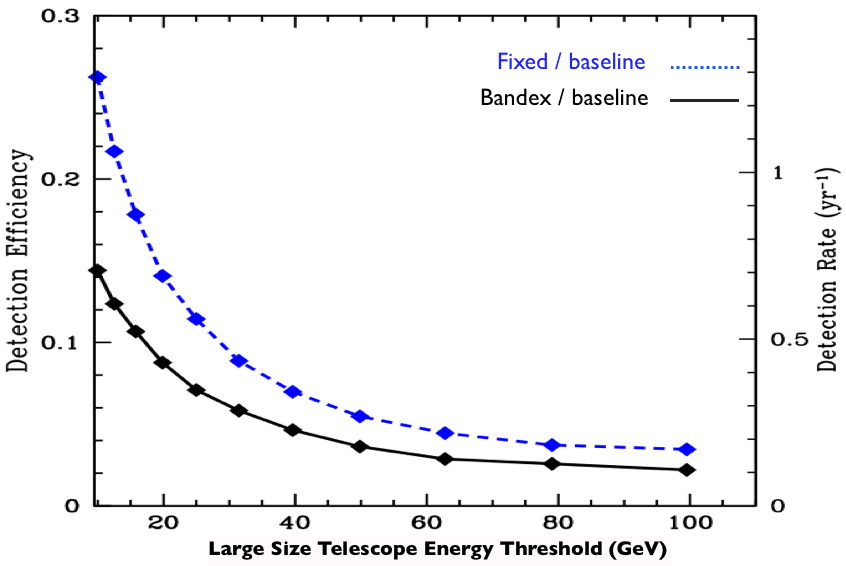}
  \caption{Detection rate as a function of the energy threshold of the Large Size Telescopes. Our standard CTA baseline and optimistic performance assume 25~GeV and 10~GeV respectively.}
  \label{Eth}
 \end{figure}

\begin{figure}[!t]
  \vspace{5mm}
  \centering
  \includegraphics[width=3.in]{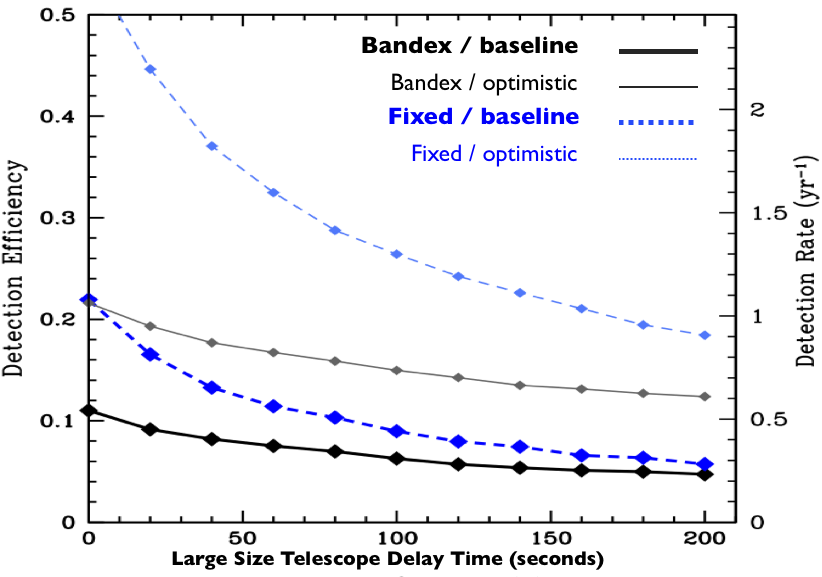}
  \caption{Detection rate as a function of the time delay for the Large Size Telescopes to reach the field of the GRB. Our standard CTA model assumes 60 seconds.}
  \label{Tdelay}
 \end{figure}

Figures \ref{Eth} and \ref{Tdelay} show how the predicted detection rates change with the two most critical parameters that we find: the LST energy threshold and time delay. Not too surprisingly, the energy threshold is crucial to catch the GRB emission which is softened by absorption through interaction with the EBL. Our study shows that a decrease in energy threshold from $\sim 25$~GeV to $\sim 10$~GeV threshold would improve the GRB detection rate by a factor of $\sim 2$.
The time delay for starting data taking has also a significant impact (fig. \ref{Tdelay}). The CTA telescope slewing times that are aimed at for the design (LSTs: $\sim 20$ sec; MSTs: $\sim 60$ s) are already quite ambitious. A crucial aspect though will be to make sure that the other two factors contributing to the overall time delay (delay of the GCN alert to be received and time for the observers to authorize slewing) are as low as possible at the time CTA starts operation.
 
We note that moonlight observations will probably be routine for CTA therefore increasing the duty cycle from $\sim~10\%$ to $\sim~13\%$. However this does not translate quite linearly into a GRB detection rate increase since the energy threshold of such observations will be substantially higher.

\subsection{Optimizing CTA observing strategy for GBM alerts}
\label{GBM}

\begin{figure}[!t]
  \vspace{5mm}
  \centering
  \includegraphics[width=3.in]{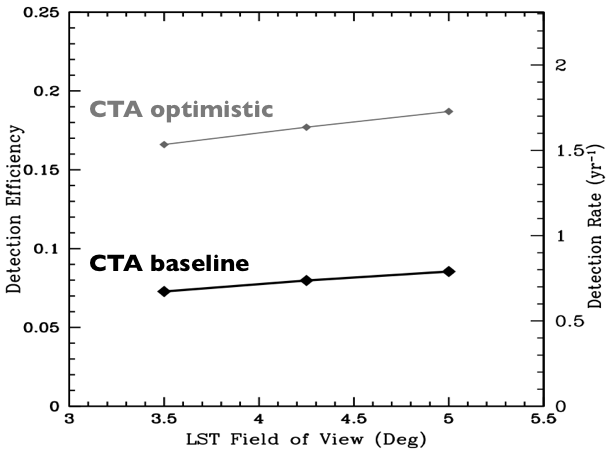}
  \caption{Detection rate as a function of the field-of-view of the Large Size Telescopes.}
  \label{GBMrates}
 \end{figure}

GBM alerts are a factor $\sim 2.5$ more frequent than $\it Swift$ alerts. However the statistical and systematic uncertainties on the burst localization are on order of several degrees, and in most cases the error box will be substantially larger than the LST Field-of-View (FoV) targeted around $4^{\circ}$ diameter. The simplest strategy for observing GBM bursts would be to point the CTA array at the best GBM localization and hope for the burst position to fall within the CTA FoV. However this results in extremely small detection rates: $\sim 0.05$ GRB/year. A smarter strategy would be to implement a so-called `orbit' mode where the GBM error box is being continuously scanned by the LST FoV.  Predicted detection rates for such `orbit' mode are significantly more promising reaching a similar level as the detection rates from {\it Swift} alerts. Figure \ref{GBMrates} shows the predicted detection rates for different assumptions on the LST FoV.  We point out that for the `orbit' mode, CTA will only be observing the field of the GRB a fraction of the time, therefore only providing useful data on small chunks of the GRB light curve. A partial solution would be to achieve real-time detection and localization of the burst to subsequently point permanently on the GRB field.

Figure \ref{GBMrates} shows a very small improvement in detection rates when the LST FoV is varied which is due to the fact that the GBM error box is typically much larger. Therefore, a further increase in detection rate (as well as light curve coverage) of GBM bursts would necessitate a significant reduction of the GBM localization errors. 
Another observing strategy has been proposed to cover the GBM error box: split the CTA array in multiple sub-arrays in order to achieve a uniform spatial and temporal coverage of the search region (which would come at the cost of lower sensitivity). However, this would only be possible with the MSTs since the LSTs are not numerous enough to cover the GBM error box. The energy threshold of MSTs being significantly higher, this observing mode would certainly lead to a much lower GRB detection rate than the `orbit' mode which is therefore highly preferred by our study.

\clearpage


\small

\begin{thebibliography}{}

\bibitem{gehrels09}
N.~{Gehrels}, E.~{Ramirez-Ruiz}, and D.~B. {Fox}, \emph{Gamma-Ray Bursts in the Swift Era}, \emph{ARA\&A} 47,
  567--617 (2009)
  
\bibitem{SVOM} Paul, J., Wei, J., Basa, 
S., \& Zhang, S.-N.\ 2011, Comptes Rendus Physique, 12, 298 

\bibitem{GBM} Meegan, C., et al. 2009, ApJ, 702, 791 

\bibitem{ghisellini10}
G.~{Ghisellini}, G.~{Ghirlanda}, L.~{Nava}, and A.~{Celotti}, \emph{GeV emission from gamma-ray bursts: a radiative fireball?}, \emph{MNRAS}
  403, 926--937 (2010)
  
\bibitem{Gilmore2011} Gilmore, R.~C., 
Somerville, R.~S., Primack, J.~R., 
\& Dom{\'{\i}}nguez, A.\ 2011, arXiv:1104.0671 
  
\bibitem{Gilmore2010} Gilmore, R.~C., Prada, 
F., \& Primack, J.\ 2010, MNRAS, 402, 565 

\end{thebibliography}
\end{document}